# Ultrahigh Charge-to-Spin Conversion and Tunneling Magnetoresistance in Quasi-Two-Dimensional *d*-wave Altermagnet

Qing Zhang[1], Siyun Wang[1], Jianting Dong[1], and Jia Zhang[1*]

*1School of Physics and Wuhan National High Magnetic Field Center, Huazhong University of Science and Technology, 430074 Wuhan, China.*

*jiazhang@hust.edu.cn

## Abstract

The emergence of altermagnets has driven groundbreaking advances in spintronics. Notably, d-wave altermagnets support non-relativistic spin transport, efficient charge-to-spin conversion, and T-odd spin currents. In addition, their integration as electrodes in antiferromagnetic tunnel junctions (AFMTJs) enables a tunneling magnetoresistance (TMR) effect, allowing electrical detection of Néel vectors for next-generation memory devices. In this work, we investigate the non-relativistic spin transport properties of the quasi-two-dimensional (quasi-2D) *d*-wave altermagnet $KV_2Se_2O$ and the TMR effect in $KV_2Se_2O$-based AFMTJs via first-principles calculations. Our results reveal that $KV_2Se_2O$ exhibits a non-relativistic longitudinal spin polarization and a spin Hall angle both exceeding 60% at room temperature, while $KV_2Se_2O$-based AFMTJs achieve a giant TMR reaching approximately $10^{12}$%, which remains robust against Fermi level shifts. These findings highlight the anisotropic spin polarization inherent to *d*-wave staggered magnetism and underscore the critical role of Fermi surface topology in enhancing *T*-odd spin transport and the TMR effect in AFMTJs.

## Introduction

The *T*-odd spin currents are pivotal for spintronics applications, owing to their intimate correlation with key performance metrics such as the non-relativistic spin Hall effect(SHE), spin torque[1][2][3][4] and tunneling magnetoresistance (TMR). Conventional spintronic devices rely predominantly on ferromagnets, whose energy bands feature exchange spin-splitting, endowing them with the capacity to generate *T*-odd spin currents. Ferromagnet-based magnetic tunnel junctions (MTJs) typically exhibit large TMR effects(>100%)[5][6][7]. However, the high write energy consumption and stray magnetic fields inherent to ferromagnets pose critical technical bottlenecks for advancing the performance of magnetic random-access memory (MRAM)[8].

These challenges in ferromagnetic MRAM hold promise for addressing via antiferromagnetic tunnel junctions (AFMTJs), as antiferromagnets possess zero net magnetization, ultrafast spin dynamics, and robustness against external magnetic field. However, the reorientation of the Néel vector in spin-degenerate antiferromagnets fails to yield significant TMR in AFMTJs. Notably, recent studies have demonstrated that a class of spin-split antiferromagnets with breaking $PT\tau$ and $U\tau$ symmetries (where $P$ denotes spatial inversion, $T$ time reversal, $\tau$ fractional translation, and $U$ spinor symmetry operations), exhibits momentum-dependent spin splitting even in the absence of spin-orbit coupling (SOC)[9][10].

Spin-splitting antiferromagnets encompass a diverse class of collinear and non-collinear materials[11]. Recent theoretical studies have predicted that the AFMTJs based on collinear[12][13] and noncollinear[14][15][16] spin-splitting antiferromagnets could achieve TMR values on the order of several hundred percent. For instance, $Mn_3Pt$- and $Mn_3Sn$-based AFMTJs have been shown to exhibit room-temperature TMR of ~100%[16] and ~2%[15], respectively. However, the TMR of AFMTJs remain significantly lower than that of ferromagnetic MTJs, such as the prototypical CoFeB/MgO/CoFeB structures, which exhibit ~200% TMR at room temperature. Practical MRAMs applications require a TMR of approximately 150%, highlighting the pressing need to boost TMR performance in antiferromagnetic tunnel junctions.

In collinear spin-splitting antiferromagnets, the opposite spin sublattices are connected by rotation symmetry, known as altermagnets. Based on the symmetry of spin-polarization in momentum space, altermagnets are classified into *d*-wave, *g*-wave, and *i*-wave types[17][18]. For several prototypical altermagnets (e.g., *d*-wave RuO₂ and *g*-wave CrSb), their spin transport properties have been experimentally validated [17, 18, 19-29]. Notably, the antiferromagnetic order in RuO₂ remains controversial[29][30][31], while CrSb, owing to its *g*-wave character only generates weakly spin-polarized currents.

Due to the rotational symmetry of spin-resolved Fermi surfaces, altermagnets exhibit intrinsic anisotropic spin polarization[32]. Compared to *g*-wave and *i*-wave counterparts, *d*-wave altermagnets (e.g. KV₂Se₂O) exhibit remarkably anisotropic transport properties[32]. This anisotropy offers design flexibility for spintronic devices and expands the application potential of altermagnets in spintronics[32][33].

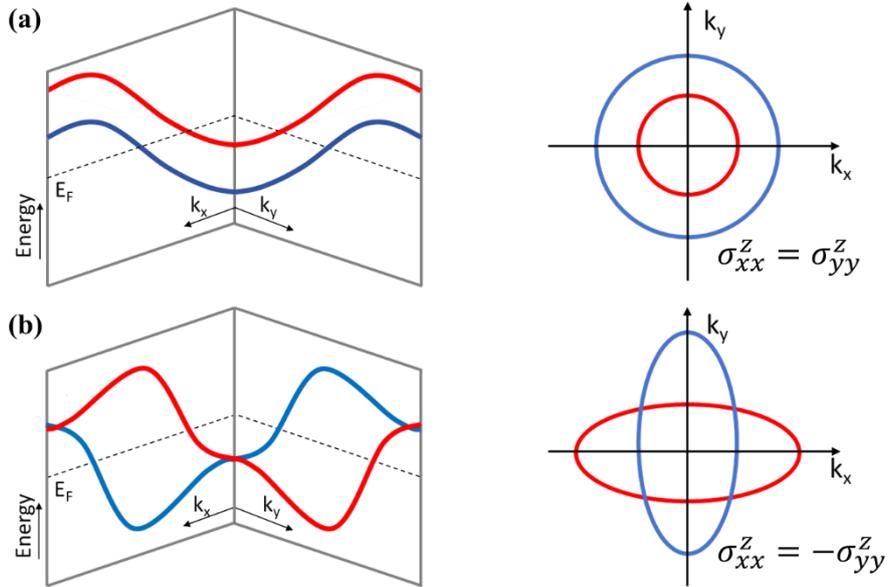

Fig. 1. Schematic band structures and Fermi surfaces of (a) isotropic ferromagnets and (b) anisotropic *d*-wave altermagnets.

As illustrated in Fig. 1, unlike ferromagnets, the spin-split band structures and spin currents of *d*-wave altermagnets are highly anisotropic and dictated by rotational symmetry. It is noteworthy that a recent discovered metallic *d*-wave altermagnet, KV₂Se₂O, exhibits a magnetic transition temperature well above room temperature [34][35]. Owing to the anisotropy of its $C_2$ symmetric spin-polarized Fermi surface, the group velocities of the two spin channels show a striking difference. As depicted in

Fig.1 (b), anisotropic spin currents and the associated non-relativistic SHE are expected to arise in the *d*-wave altermagnet KV$_2$Se$_2$O. Furthermore, compared to their three-dimensional counterparts, the overlap between the spin-up and spin-down Fermi surfaces should be extremely narrow in quasi-2D altermagnets. Consequently, AFMTJs based on such altermagnets hold great promise for realizing giant TMR.

In this work, the spin conductivity of KV$_2$Se$_2$O and TMR in corresponding antiferromagnetic tunnel junctions (AFTJs) have been investigated via first-principles calculations. Our results reveal significant longitudinal spin polarization—even exceeding that of ferromagnets. A giant non-relativistic spin Hall angle, originating from the anisotropic longitudinal spin conductivity, has been predicted. Furthermore, we demonstrate a record-high tunnel magnetoresistance (TMR) in KV$_2$Se$_2$O-based AFMTJs, which exhibits remarkable robustness to Fermi energy shifts.

## Computational methods

The spin conductivity for KV$_2$Se$_2$O was calculated via the scalar-relativistic multiple-scattering Korringa-Kohn-Rostoker (KKR) Green's function method, based on the Kubo-Bastin linear response formalism [36][37]. Phonon scattering at 300 K was incorporated by means of the alloy analogy model.

The transport properties of KV$_2$Se$_2$O-based AFMTJs were calculated using the Quantum ESPRESSO package[38], which employs a plane-wave basis set. The exchange-correlation functional was treated within the generalized gradient approximation (GGA) of the Perdew-Burke-Ernzerhof (PBE) form[39], alongside the ultrasoft pseudopotential approach[40]. The electronic structure of KV$_2$Se$_2$O was computed self-consistently with lattice parameter fixed at experimental values (*a*=*b*=3.946Å and *c*=7.312Å), using a 15×15×8 Monkhorst ***k***-points mesh for Brillouin Zone sampling. The plane-wave and charge density cutoff energies were set to 40 Ry and 480 Ry respectively.

Cubic perovskite SrTiO$_3$, with lattice constant *a*=*b*=*c*=3.913 Å and an extremely small lattice mismatch with KV$_2$Se$_2$O, has been selected as the tunnel barrier. The electronic structure of the KV$_2$Se$_2$O|SrTiO$_3$|KV$_2$Se$_2$O interface was calculated self-consistently

using a *k*-point mesh of 15×15×1. Then the electron transmission was computed by the wave function scattering method[41] with a 400×400 $k_\parallel$ mesh over the two-dimensional Brillouin zone (2DBZ). The spin dependent conductance is obtained by summing transmission probabilities over the entire 2DBZ:

$$G_\sigma = \frac{e^2}{h} \sum_{k_\parallel} T_\sigma(k_\parallel) \tag{2}$$

where $T_\sigma(k_\parallel)$ is the transmission probabilities at Fermi energy with spin index $\sigma$ and the Bloch wave vector $k_\parallel(k_x, k_y)$, $e$ is the elementary electron charge and $h$ is the Plank constant.

## Results and discussions

Given our focus on non-relativistic spin transport phenomena, the spin space group (SSG) is well-suited for analyzing the symmetry of $KV_2Se_2O$. Notably, preserved symmetry relations between the two spin sublattices (e.g., glide mirrors or screw axes) impose additional constraints on the band structures, leading to distinct spin-dependent characteristics in momentum space. As depicted in Fig. 2 (a), $KV_2Se_2O$ features a layered tetragonal crystal structure consisting of $V_2Se_2O$ layers separated by K atomic layer. The collinear antiferromagnetic (AFM) phase of $KV_2Se_2O$ belongs to the SSG $P^{-1}4/^1m^1m^{-1}m^{\infty m}1$, with the opposite-spin sublattices related by by the $[C_2||C_{4z}]$ symmetry operation.

The calculated spin-polarized band structure and Fermi surfaces of bulk $KV_2Se_2O$ are presented in Fig.2 (b)-(c). Unlike the curved, closed surfaces of conventional bulk materials, that of $KV_2Se_2O$ exhibits a nearly flat geometric morphology along specific *k*-paths in the Brillouin Zone. This distinctive feature indicates that $KV_2Se_2O$ may exhibit extremely anisotropic spin transport behavior, which facilitates the spatial separation of spin channels and the emergence of the non-relativistic spin Hall effect. Previous studies have demonstrated that such flat Fermi surfaces can generate intensive spin currents[42]. The band structure displays spin splitting along the Γ-X-M and Γ-Y-M *k*-paths, while spin degeneracy is maintained along the Γ-M *k*-path, which is in excellent agreement with prior computational results. Notably, the symmetry of band structure and Fermi surface in momentum space is perfectly consistent with the $[C_2\|C_{4z}]$ symmetry operation in real space.

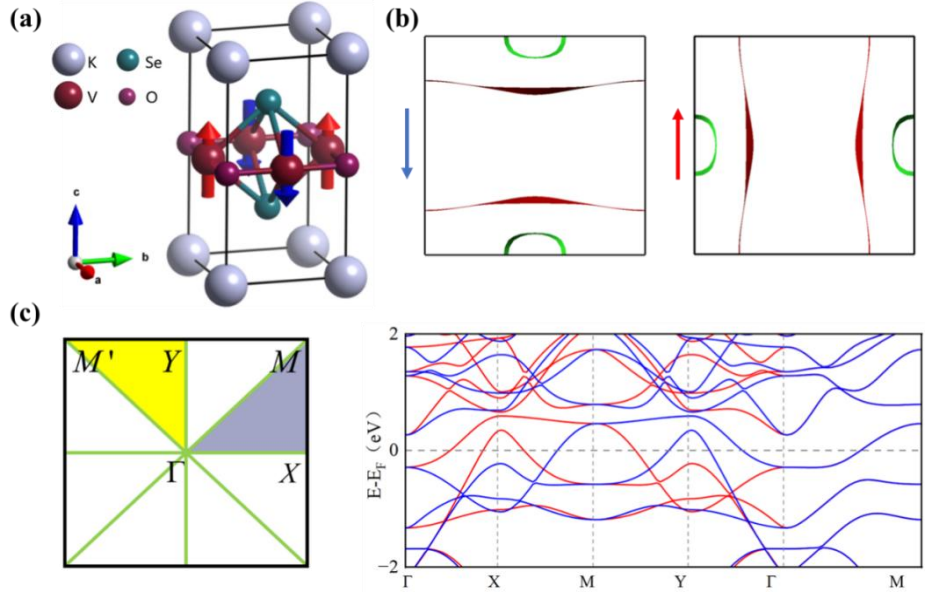

Fig. 2. (a) Crystal and magnetic structures of $KV_2Se_2O$, with red and blue arrows indicating the magnetic moments carried by V atoms.(b) Ab initio calculated spin-up and spin-down Fermi surfaces of bulk $KV_2Se_2O$. (c) High-symmetry ***k***-path in the first Brillouin zone and corresponding spin-resolved band structure of $KV_2Se_2O$; red and blue lines stand for spin-up and spin-down bands, respectively.

The linear response spin current driven by an electric field is described by the following expression:

$$J_\mu^s = \sum_\nu J_{\mu\nu}^s = \sum_\nu \sigma_{\mu\nu}^s E_\nu, (\mu,\nu,s \in \{x,y,z\}) \qquad (1)$$

where $\nu$ denotes the electric field direction, $J_\mu^s$ represents the spin current propagating along the $\mu$ direction with spin polarization $s$, and $\sigma_{\mu\nu}^s$ are the corresponding spin conductivity tensor elements. The form of the spin conductivity tensor is constrained by the symmetry operations if the point group. $KV_2Se_2O$ with spin quantum axis along $z$ exhibits two non-relativistic longitudinal spin conductivity components, $\sigma_{xx}^z$ and $\sigma_{yy}^z = -\sigma_{xx}^z$, as dictated by the $[C_2\|C_4z]$ spin point group symmetry. Conversely, the symmetry operations $[m_{\perp 2}\|1]$ and $[m_{\|}\|1]$ enforce vanishing spin conductivity for *x*- and *y*-polarized spins. Our calculation results indicate that only two

non-zero spin conductivity tensor elements exist, with a magnitude of $\sigma^z_{xx} = -\sigma^z_{yy} = -2.43 \times 10^5$ $\hbar/2e$(S/m), while all other elements are zero. Additionally, the charge conductivity tensors of KV$_2$Se$_2$O exhibits longitudinal components $\sigma_{xx} = \sigma_{yy} = 4.02 \times 10^5$(S/m) and $\sigma_{zz} = 1.57 \times 10^4$(S/m). As shown in Fig. 3, the anisotropic spin conductivity of KV$_2$Se$_2$O give rise to a non-relativistic SHE, whose magnitude shows anisotropy as a function of the electric field direction. Based on the current density formula, if the angle between the electric field and the original *x*-axis is $\varphi$, the spin Hall and longitudinal spin conductivity as functions of $\varphi$ are found to be $\sigma^z_{x'y'} = \sigma^z_{y'x'} = -\sigma^z_{xx}\sin(2\varphi)$ and $\sigma^z_{x'x'} = -\sigma^z_{y'y'} = \sigma^z_{xx}\cos(2\varphi)$, respectively.

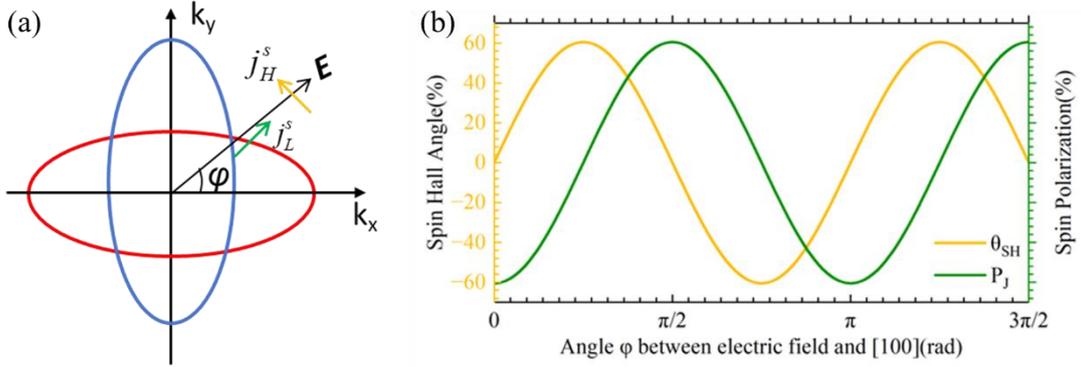

Fig. 3 (a) Schematic illustration of longitudinal and Hall spin current configurations in *d*-wave KV$_2$Se$_2$O. (b) Calculated Spin Hall angle (yellow line, left axis) and longitudinal spin polarization (green line, right aixs) of KV$_2$Se$_2$O versus the angle between the applied electric field and the [100] crystallographic direction.

The spin Hall angle (SHA) and longitudinal spin-polarization (SP) are defined as: SHA=$\sigma^z_{x'y'}/\sigma_{xx}$, SP=$\sigma^z_{x'x'}/\sigma_{xx}$. We find that the maximum longitudinal spin-polarization reaches 60%, even exceeding that of typical ferromagnetic metals. Furthermore, the spin Hall angle surpasses that driven by the SOC mechanism by nearly one order of magnitude. These findings highlight that *d*-wave KV$_2$Se$_2$O with exceptional spin transport properties should be useful for applications, such as the giant spin split torque in current induced switching in "KV$_2$Se$_2$O/ferromagnet" bilayer, where KV$_2$Se$_2$O serves as the spin current source[26].

Although the net longitudinal spin-polarization of KV$_2$Se$_2$O is zero when the electric field is along *z* direction, i.e. $\sigma^z_{zz} = 0$, there is spin splitting in momentum space due to the breaking $PT\tau$ and $U\tau$ symmetries. We thus establish and investigate the TMR effect in $KV_2Se_2O/SrTiO_3/KV_2Se_2O$ tunnel junctions. As illustrated in Supplementary Fig. S1, four distinct $KV_2Se_2O/SrTiO_3$ interface configurations exist. A lower formation energy implies that the interface structure is thermodynamically more stable. Subsequently, we calculated the interface formation energy between $SrTiO_3$ and $KV_2Se_2O$, and ultimately identified that the SrO-K interface as the most stable configuration (with the lowest formation energy). The interfacial separation was determined to be 3.17 Å.

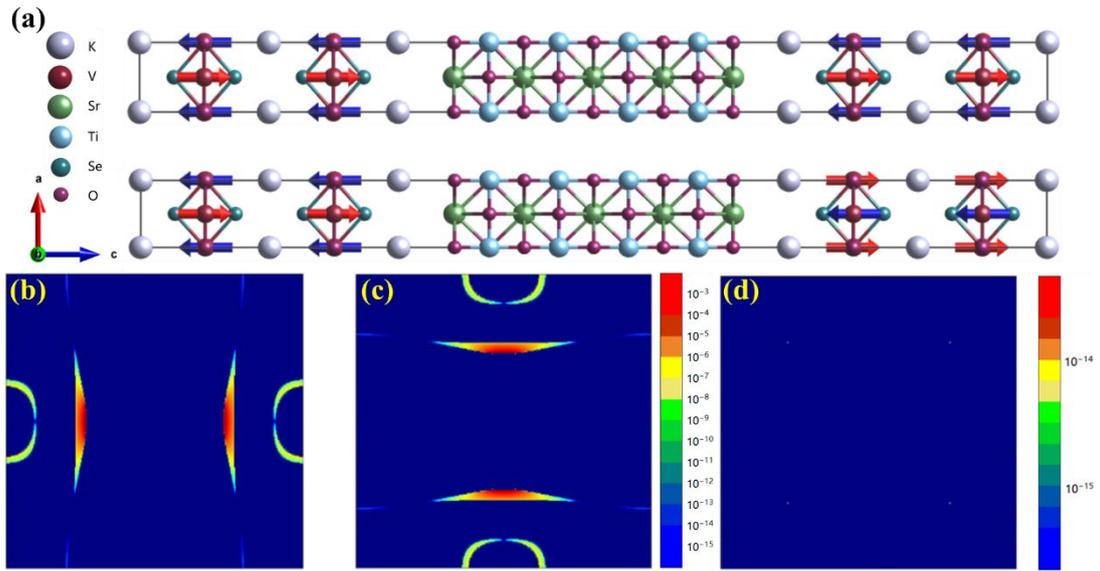

Fig. 4. Atomic structure and spin-resolved transmission of the tunnel junction in parallel (P) and antiparallel (AP) magnetic configurations . (a) Atomic structure of the tunnel junction. (b) Spin-up transmission in the P state. (c) Spin-down transmission in the P state.(d) Spin-up transmission in the AP state.

In AFMTJs, the insulating tunnel barrier significantly modulates transport properties. The wave function decay rate within the barrier exhibits momentum dependence. Consequently, the distribution of the barrier's tunneling decay rate in reciprocal space affects the electron tunneling probability and its dependence on in-pane $k$ vectors ($k_{\parallel}$ ($k_x$, $k_y$)) , which in turn tunes the TMR effect of the junction. For $KV_2Se_2O$, if the maximum decay rate within the barrier coincides with the overlapping regions of the Fermi surfaces from the two spin channels, this configuration can effectively suppress tunneling in the anti-parallel state, thereby yielding an enhanced TMR effect. To elucidate the correlation bewteen transport properties and the Fermi surface topopgy of KV$_2$Se$_2$O, we selected SrTiO$_3$ with C$_{4z}$ symmetry, where $z$ denotes the junction's transport direction. Notably, the $k_\parallel$-dependent decay rate distribution of SrTiO$_3$ also exhibits fourfold rotational symmetry [43].

The calculated spin- and $k_\parallel$-resolved transmission for the parallel and antiparallel magnetization configurations of the MTJ are presented in Fig. 4(b)-(d). For the parallel configuration, the transmission pattern is found to resemble the corresponding spin-up and spin-down Fermi surface topologies. In contrast, for the antiparallel configuration, near-complete mismatch between the Fermi surfaces of the spin-up and spin-down channels give rise to a pronounced suppression of transmission. Specially, the computed total transmission values for the parallel and antiparallel configurations are $3.33\times10^{-5}$ and $4.06\times10^{-17}$ respectively, corresponding to a giant TMR ratio on the order of $10^{12}$%. The TMR effect is highly dependent on the Fermi surface topology. However, practical materials are subject to inevitable perturbations including doping, strain, and interfacial effects—all of which can induce a Fermi energy shift. We therefore evaluated the TMR as a function of energy over a range of -0.2 eV to 0.2 eV relative to the Fermi level. The corresponding TMR versus Fermi energy shift is plotted in Supplementary Fig. S4. As depicted, the TMR of the KV$_2$Se$_2$O/SrTiO$_3$/KV$_2$Se$_2$O tunnel junction remains robustly above $10^{10}$% even when the Fermi energy varies within the range of -0.2 eV to 0.2 eV. Futhermore, within this energy window, the TMR effect exhibits a monotonic upward trend with increasing Fermi level. Especially, it peaks at a value of $10^{16}$% when the Fermi level shifts to 0.2 eV.

These observations are further corroborated by analyzing the spin-resolved Fermi surfaces at different Fermi energies (Supplementary Fig.S2 ). When the Fermi level varies within the range of -0.2 eV to 0.2 eV, the overlapping regions of the spin-resolved Fermi surfaces remain narrow. The robust TMR effect exhibited by KV$_2$Se$_2$O highlights the critical dependence of the TMR effect on Fermi surface topology. Notably, KV$_2$Se$_2$O emerges as a highly promising candidate for practical applications in AFMTJ-based spintronic devices, as it exhibits a record-high TMR compared to other reported AFMTJs (see Supplementary Note 4).

## Conclusion

This study focuses on the quasi-two-dimensional *d*-wave altermagnet KV$_2$Se$_2$O. By combining symmetry analysis and first-principles calculations, we systematically its transport properties, verifying that the spin conductivity tensor conforms to theoretical projections. Using KV$_2$Se$_2$O as electrode materials, we constructed an antiferromagnetic tunnel junction (AFMTJ) with a KV$_2$Se$_2$O|SrTiO$_3$|KV$_2$Se$_2$O heterostructure.

First-principles calculations demonstrate that KV$_2$Se$_2$O exhibit a spin polarization ratio and spin Hall angle both exceeding 60% at room temperature (300 K). Notably, KV$_2$Se$_2$O-based AFMTJs are predicted to yield an exceptionally high tunneling magnetoresistance (TMR) ratio. Crucially, this giant TMR effect remains robust against Fermi energy shifts. We anticipate that these conclusions can be extended to analogous materials of the XV$_2$Y$_2$O family(X = K, Rb; Y = S, Se)[44], which hold substantial potential for applications in AFMTJs and other advanced spintronic devices.

## Acknowledgement

This work was supported by the National Natural Science Foundation of China (Grant No. T2394475, T2394470, 12174129).